\documentclass[11pt,preprint]{aastex}
\begin{document}

\title{Observations of the Magnetic Cataclysmic Variable VV Puppis 
with the {\em Far~Ultraviolet Spectroscopic Explorer}\footnote{Based 
on observations with the NASA-CNES-CSA {\em Far Ultraviolet Spectroscopic 
Explorer}. {\em FUSE} is operated for NASA by the Johns Hopkins 
University under NASA contract NAS 5-32985.}}

\author{D.\ W.\ Hoard and Paula Szkody}
\affil{\vspace*{-11pt}Department of Astronomy, University of Washington, 
Box 351580, Seattle WA 98195-1580} 
\email{\vspace*{-11pt}hoard, szkody@astro.washington.edu}
\authoraddr{Department of Astronomy, University of Washington, Box 351580, 
Seattle WA 98195-1580} 

\author{Ryoko Ishioka}
\affil{\vspace*{-11pt}Department of Astronomy, Faculty of Science, 
Kyoto University, Sakyou-ku, Kyoto, 606-8502, Japan}
\email{\vspace*{-11pt}ishioka@kusastro.kyoto-u.ac.jp}

\author{L.\ Ferrario}
\affil{\vspace*{-11pt}Department of Mathematics, Australian National 
University, Canberra, ACT0200, Australia} 
\email{\vspace*{-11pt}Lilia.Ferrario@maths.anu.edu.au}

\author{B.\ T.\ G\"{a}nsicke}
\affil{\vspace*{-11pt}Department of Physics and Astronomy, University 
of Southampton, Highfield, Southampton SO17~1BJ, UK} 
\email{\vspace*{-11pt}btg@astro.soton.ac.uk}

\author{Gary D.\ Schmidt}
\affil{\vspace*{-11pt}Steward Observatory, University of Arizona, 
933 North Cherry Avenue, Tucson, AZ 85721-0065} 
\email{\vspace*{-11pt}gschmidt@as.arizona.edu}

\author{Taichi Kato and Makoto Uemura}
\affil{\vspace*{-11pt}Department of Astronomy, Faculty of Science, 
Kyoto University, Sakyou-ku, Kyoto, 606-8502, Japan}
\email{\vspace*{-11pt}tkato, uemura@kusastro.kyoto-u.ac.jp}

\begin{abstract}
We present the first far-ultraviolet (FUV) observations of the 
magnetic cataclysmic variable VV Puppis, obtained with the 
{\em Far Ultraviolet Spectroscopic Explorer} satellite.  
In addition, we have obtained simultaneous ground-based optical 
photometric observations of VV Pup during part of the FUV observation. 
The shapes of the FUV and optical light curves are consistent with 
each other and with those of past observations at optical, 
extreme-ultraviolet, and X-ray wavelengths.  Time-resolved FUV 
spectra during the portion of VV Pup's orbit when the accreting 
magnetic pole of the white dwarf can be seen show an increasing 
continuum level as the accretion spot becomes more directly visible.  
The most prominent features in the spectrum are the 
\ion{O}{6} $\lambda\lambda$1031.9, 1037.6 emission lines.  
We interpret the shape and velocity shift of these lines in the 
context of an origin in the accretion funnel near the white dwarf 
surface.  A blackbody function with $T_{\rm bb}\gtrsim90,000$ K 
provides an adequate fit to the FUV spectral energy distribution 
of VV Pup.
\end{abstract}


\slugcomment{Accepted for publication in the Astronomical Journal, 06/25/02}

\section{Introduction}

VV Puppis was noticed early in the 20th century \citep{vangent31} 
as a faint \cite[$V=14.5$--$18$;][]{downes01}, rapidly periodic 
($P\approx100$ min) variable.  Based on photometric and 
spectroscopic observations, \citet{herbig60} suggested it was a 
binary whose emission lines originate on the brighter component.  
Later, VV Pup was identified as the third member of the AM Herculis 
class of magnetic cataclysmic variable \cite[CV;][]{tapia77}.  
The white dwarf (WD) primary star in an AM Her system has a magnetic 
field strength of $B \gtrsim 10$ MG (currently known to reach up to 
several hundred MG; e.g., $B = 230$ MG in AR Ursae Majoris, 
\citealt{schmidt99}).  The magnetic field prevents the formation of 
the accretion disk that dominates the luminosity of non-magnetic CVs.  
Instead, the accretion stream emerging from the low mass, main 
sequence secondary star's Roche lobe is entrained onto the WD's 
magnetic field lines and funneled directly onto its magnetic pole(s) 
\cite[see review in][ch.\ 6]{warner95}.  
Other distinguishing characteristics of the AM Her stars include WD 
spin periods synchronized with their orbital periods and a high 
degree of linear and circular polarization (leading to the alternate 
name ``polars'' for this class of CV).
Occasional reductions or interruptions to the accretion flow cause 
AM Her systems to drop to ``low states'' of reduced brightness.  
The origin of this accretion modulation is not fully known but may 
be related to, for example, solar-type magnetic activity (starspots) 
on the secondary star \citep{hessman00}.

Modeling of the cyclotron lines in the optical spectrum of VV Pup 
(which were first noted by \citealt{visv79} and \citealt{wick79}) 
allowed \citet{barrett85} to estimate a magnetic field strength of 
$B=31.5$ MG at the accreting pole.  The pole is located at an 
azimuth (i.e., the angle between the line of centers of the component 
stars and the projection of the magnetic axis onto the orbital 
plane of the CV) of $\psi\approx50^{\circ}$ and a colatitude 
(i.e., the angle between the rotation and magnetic axes of the WD; 
see \citealt{warner95}, his Figure 6.3, for definitions of angles 
in polar geometry) of $\delta\approx150^{\circ}$.  
Combined with a system inclination of 
$i\approx75^{\circ}$ for VV Pup, this pole is visible for 
$\approx$45\% of the CV's orbit (\citealt{cropper88} and 
references therein).  
\citet{wick89} later used observations obtained during a very high
accretion state to detect a 
$B=56$ MG field at the second magnetic pole, which forms an off-center 
dipole with the weaker first pole.
Despite being weaker, the location of the first pole on the 
synchronously rotating WD -- on the side facing the secondary 
star -- makes it the preferred site for the accretion flow.
However, accretion onto the second pole (which is located within 
$\approx10^{\circ}$ of the WD rotation axis, so is always visible; 
\citealt{wick89}) sometimes also occurs, as evidenced by 
polarimetric observations made when the first pole is not visible 
(due to orbital motion) and differences in the total system brightness 
when the first pole is inactive ($V\approx16$ when the second pole 
is active vs.\ $V\approx18$ when both poles are inactive; 
e.g., \citealt{liebert79}). A blackbody fit to the optical spectrum 
of VV Pup during an extended, steady low state (both poles inactive) 
in 1977 suggested a temperature of $\approx9000$ K for the 
WD \citep{liebert78}.

VV Pup has been extensively observed at infrared \citep{szkody83}, 
optical \citep{warner72, imamura00}, ultraviolet \citep{patterson84}, 
extreme-ultraviolet (EUV; \citealt{vennes95}), and 
X-ray \citep{patterson84, imamura00}  wavelengths.  
The EUV data indicated a possible oxygen overabundance and the 
presence of a hot accretion region.  To further study these 
properties of VV Pup, we obtained the first far-ultraviolet (FUV) 
observation of this magnetic CV, with the {\em Far Ultraviolet 
Spectroscopic Explorer} ({\em FUSE\/}) satellite.

\section{Observations}

We observed VV Pup with {\em FUSE} during 11 visits between 
HJD 2452004.5--2452005.3 (2001 April 05, 00:28--17:08 UT; 
see Table \ref{t-fuselog}).  The start time of each visit was 
selected to correspond to approximately the same orbital phase 
of the CV in successive cycles, in order to facilitate co-adding 
data spanning $\approx40$\% of the CV's orbit.
All visits utilized the LWRS aperture and TTAG accumulation 
mode (for {\em FUSE} spacecraft and instrument details see, 
e.g., \citealt{sahnow00})\footnote{Also see the {\em FUSE} Science 
Center web page at \url{http://fuse.pha.jhu.edu/}.}.
Details of the FUV data processing are given in 
Sections \ref{s-fuvlc} and \ref{s-fuvspec}.

\subsection{Far-ultraviolet Light Curves}
\label{s-fuvlc}

We used the CalFuse v2.0.5 software to extract time-resolved spectra 
from the raw data files obtained during each {\em FUSE} visit, in 
successive 200-s intervals starting at the beginning of each visit.  
Remaining exposure intervals of less than 200 s at the end of any 
visit were not used.  
We then used a custom-built IDL routine (following the recipes in 
the {\em FUSE} Data Analysis Cookbook\footnote{See 
\url{http://fuse.pha.jhu.edu/analysis/cookbook.html}.}) to combine 
the various mirror and detector segments of the spectra.  
Because the spectra were very weak, exceptions to the standard 
Cookbook procedures were made as follows:\
(1) only the LiF data ($\lambda \gtrsim 980$ \AA) were used (the 
SiC data are very noisy, with little apparent signal) and
(2) no wavelength cross-correlation or shifts between different 
exposures were applied.
There is a small gap in the wavelength coverage between 1082--1087 \AA\ 
where only SiC data are available.  
In addition, the LiF1 data for $\lambda > 1133$ \AA\ were not used in 
order to avoid the artifact known as ``The Worm'' (only the LiF2 data 
in this region are used).  The final combined spectra were rebinned 
onto a uniform wavelength scale with dispersion 0.05 \AA\ pixel$^{-1}$
by averaging flux points at the original dispersion into wavelength
bins of width 0.05 \AA.

In this manner, a total of 87 time-resolved spectra were extracted 
from the 11 {\em FUSE} visits.  
These spectra are too weak and noisy to be useful in their dispersed 
form; instead, we summed the flux values in the wavelength ranges 
(A) 1000.0--1024.5 \AA\ $+$ 1045.0--1080.0 \AA\ $+$ 1088.0--1160.0 \AA, 
and (B) 1028.5--1045.0 \AA.  
Both of these ranges were chosen to avoid the extremely noisy short 
wavelength ($\lambda < 1000$ \AA) data, a strong airglow feature at 
the location of Ly-$\beta$, and the detector gap mentioned above.  
Wavelength region A includes primarily continuum with only 
a few weak line features (and excludes region B), while region B 
includes only the 
\ion{O}{6} $\lambda\lambda$1031.9, 1037.6 lines 
(the \ion{O}{1} $\lambda\lambda$1039.2, 1040.9 lines are also 
included -- these are probably airglow features, but they are 
relatively weak compared to the \ion{O}{6} lines, so should 
contribute little to the total flux in this wavelength region).  
We henceforth refer to these data as (A) continuum and (B) emission.  

\begin{figure}[tb]
\epsscale{1.00}
\plotone{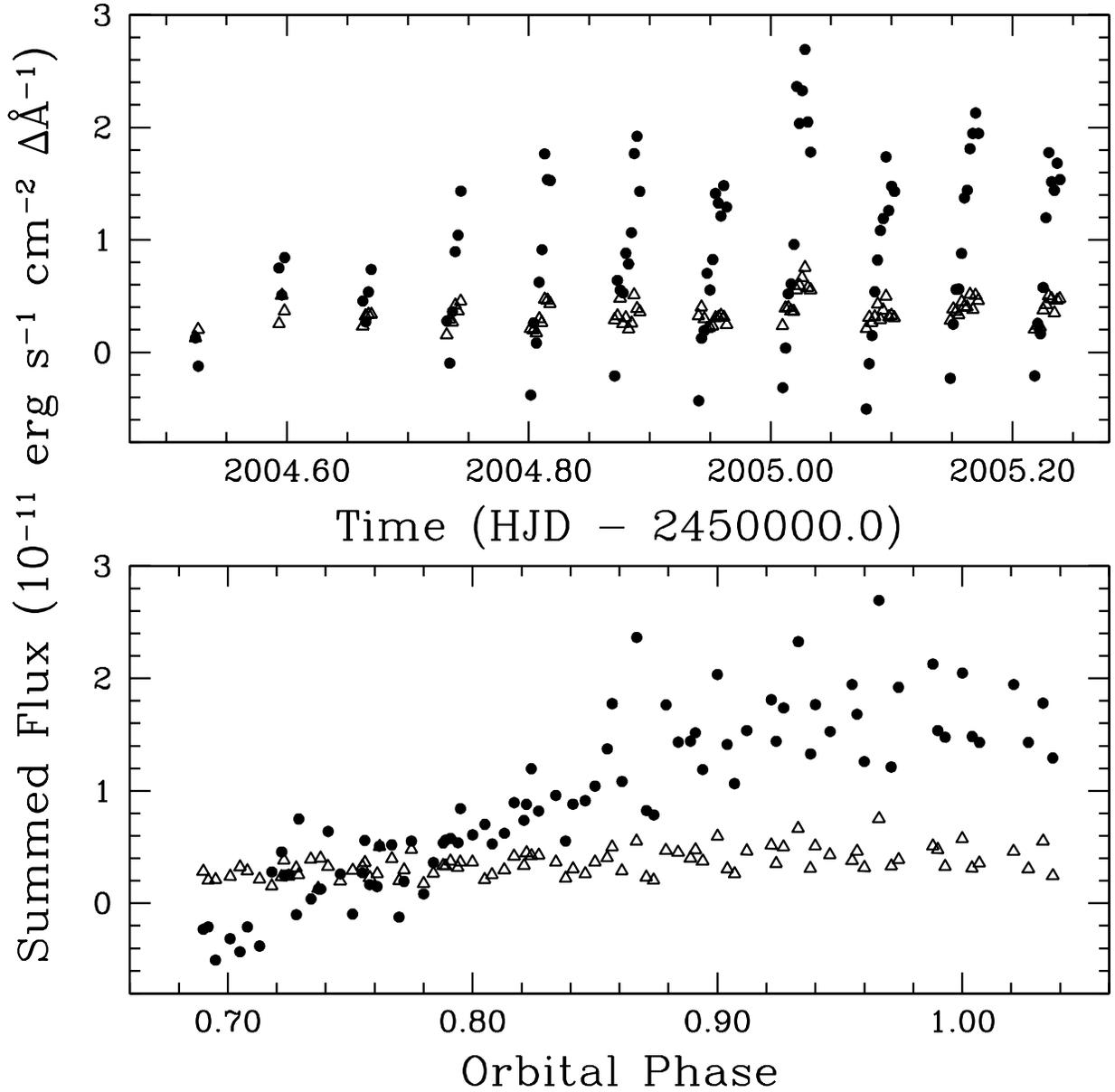}
\epsscale{1.00}
\figcaption{FUV light curves of VV Pup constructed by summing fluxes 
in 200-s time-resolved {\em FUSE} spectra (see text for wavelength 
regions summed).  The continuum fluxes are plotted as filled circles; 
the oxygen emission line fluxes are plotted as unfilled triangles.  
The top panel shows the data as a function of time, while the bottom 
panel shows the data folded on the orbital ephemeris of 
\citet{walker65}.  The 11 individual {\em FUSE} visits are apparent 
in the top panel.
\label{f-fuvlc}}
\end{figure}

Figure \ref{f-fuvlc} shows the FUV light curves constructed from 
the summed spectra.   
Orbital phases were calculated using the ephemeris of \citet{walker65}; 
we note that this ephemeris defines the phase of maximum (optical) 
light as its zero point.  Infrared observations 
(\citealt{szkody80,szkody83}; also see the geometric model 
in \citealt{patterson84}, their Figure 8) show that the inferior 
conjunction of the secondary star (normally used as the orbital 
phase zero point) occurs at $\phi\approx0.1$ according to 
the \citet{walker65} ephemeris.  This amount may be subtracted 
from the phases quoted in this paper to approximate the 
``standard'' relationship between orbital phase and system 
geometry for CVs.
The continuum flux in our {\em FUSE} light curve starts at 
zero at phase $\phi\approx0.7$, 
increases substantially to a maximum at $\phi\approx0.98$, 
then appears to begin declining at later phases.
The emission flux approximately follows the continuum pattern,
almost doubling its mean amplitude at the start of the light curve by
the time it reaches $\phi\approx0.98$.
However, the rate of increase and subsequent leveling off 
is more gradual than displayed by the continuum data.  
The continuum fluxes were summed over a wavelength region $\approx6$ 
times larger than the emission fluxes, so their absolute flux levels 
as shown in Figure \ref{f-fuvlc} cannot be directly compared to 
each other.  However, even after accounting for this difference, 
the slope of the continuum light curve (obtained by approximating 
the curve as a linear function) over the phase range 
$\phi=0.70$--$0.98$ is more than twice that of the same section 
of the emission light curve.

\subsection{Optical Light Curve}
\label{s-optlc}

\begin{figure}[tb]
\epsscale{0.85}
\plotone{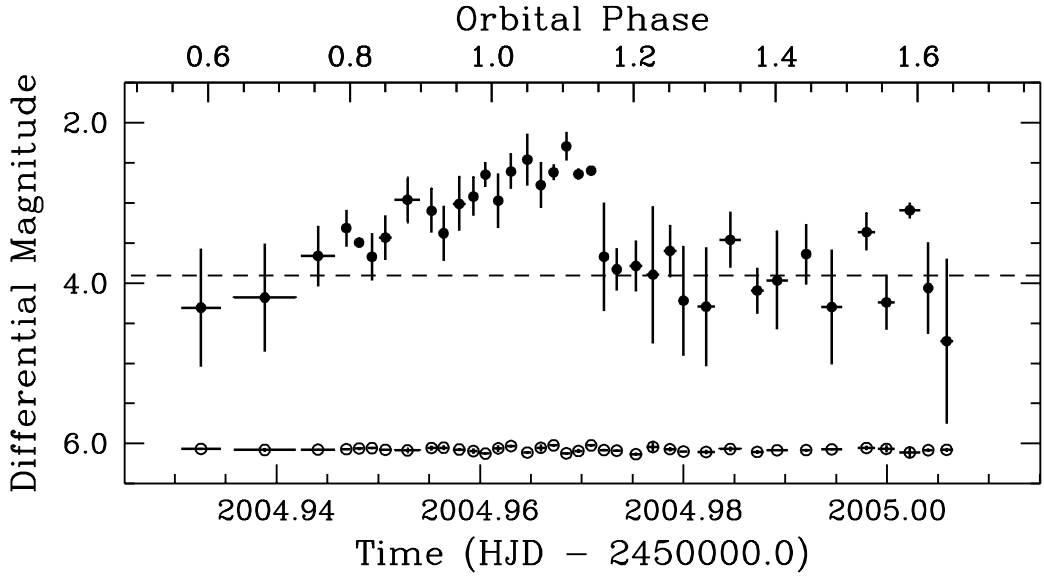}
\epsscale{1.00}
\figcaption{Optical differential light curve of VV Pup and a 
comparison star (filled circles), averaged into bins of 3 data 
points each.  The vertical error bars are the 
standard deviations of the means of each 3-point bin, while the 
horizontal bars show the width in time and phase of each bin.  
The unfilled circles are the differential light curve of another 
field star and the same comparison star (offset by +6.0 mag for 
clarity and also binned by 
3 points); $1\sigma$ vertical error bars are also plotted for 
the comparison star light curve, but they are smaller than the 
data points at the scale of the plot.  
The lower x-axis scale gives the heliocentric Julian date, while 
the upper x-axis scale gives the orbital phase of VV Pup from 
the ephemeris of \cite{walker65}.  The dashed line shows the 
effective zero level of the VV Pup light curve -- it is located 
at $y=3.903$ mag, which is the mean of the VV Pup differential 
magnitudes for HJD $>$ 2452004.9725.
\label{f-optlc}}
\end{figure}

We also obtained ground-based optical observations of VV Pup on 
2001 April 05, overlapping with {\em FUSE} visit 7 (ending just 
prior to the start of {\em FUSE} visit 8). 
These data were acquired using an 
unfiltered ST-7 CCD on a 25-cm telescope.
Each measurement had an exposure time of 30 s, and the entire time 
series spanned 0.0775 d (1.86 hr).  
We calculated instrumental magnitudes and statistical uncertainties 
from the net background-subtracted count rates in each exposure for 
VV Pup and several nearby field stars.  
Figure \ref{f-optlc} shows the differential light curve of 
VV Pup with respect to a comparison star.  
The differential light curve of another 
field star and the same comparison star is also shown to illustrate
the intrinsic noise level of the 
measurements ($\sigma_{\rm diff} = 0.046$ mag).
According to the USNO A1.0 catalog, our comparison star 
(USNO 0675-06141640) has blue and red magnitudes of 13.8 and 12.6, 
respectively.  Thus, VV Pup was fainter than $\approx16.5$ mag at 
orbital phases away from the hump in the light curve, indicating 
that the second magnetic pole was not accreting during our 
{\em FUSE} observation.
The larger scatter at each end of the VV Pup light curve is caused 
by poor- or non-detection of the CV during orbital phases when it 
is not bright (i.e., when the accreting magnetic pole has 
rotated out of view).  

The overall shape of our optical light curve is very similar to 
that of the optical light curve of 
VV Pup in \citet{patterson84} and the EUV light curve in 
\citet{vennes95}.  The latter authors note that the ephemeris 
of \citet{walker65} is good to within $\pm0.05$ orbital cycles 
at the epoch of their data (1993).  
Following this reasoning, we would expect that the phasing of our 
2001 epoch data is good to better than $\approx\pm0.07$ cycles.  
However, comparison of the phasing of the bright hump in our 
optical light curve with the published light curves listed above 
\citep[especially in][their Figure 3]{patterson84} suggests that 
the accumulated offset in the phasing of the \citet{walker65} 
ephemeris is less than $\pm0.07$, and likely still as small 
as $\approx\pm0.05$.

\begin{figure}[tb]
\epsscale{0.95}
\plotone{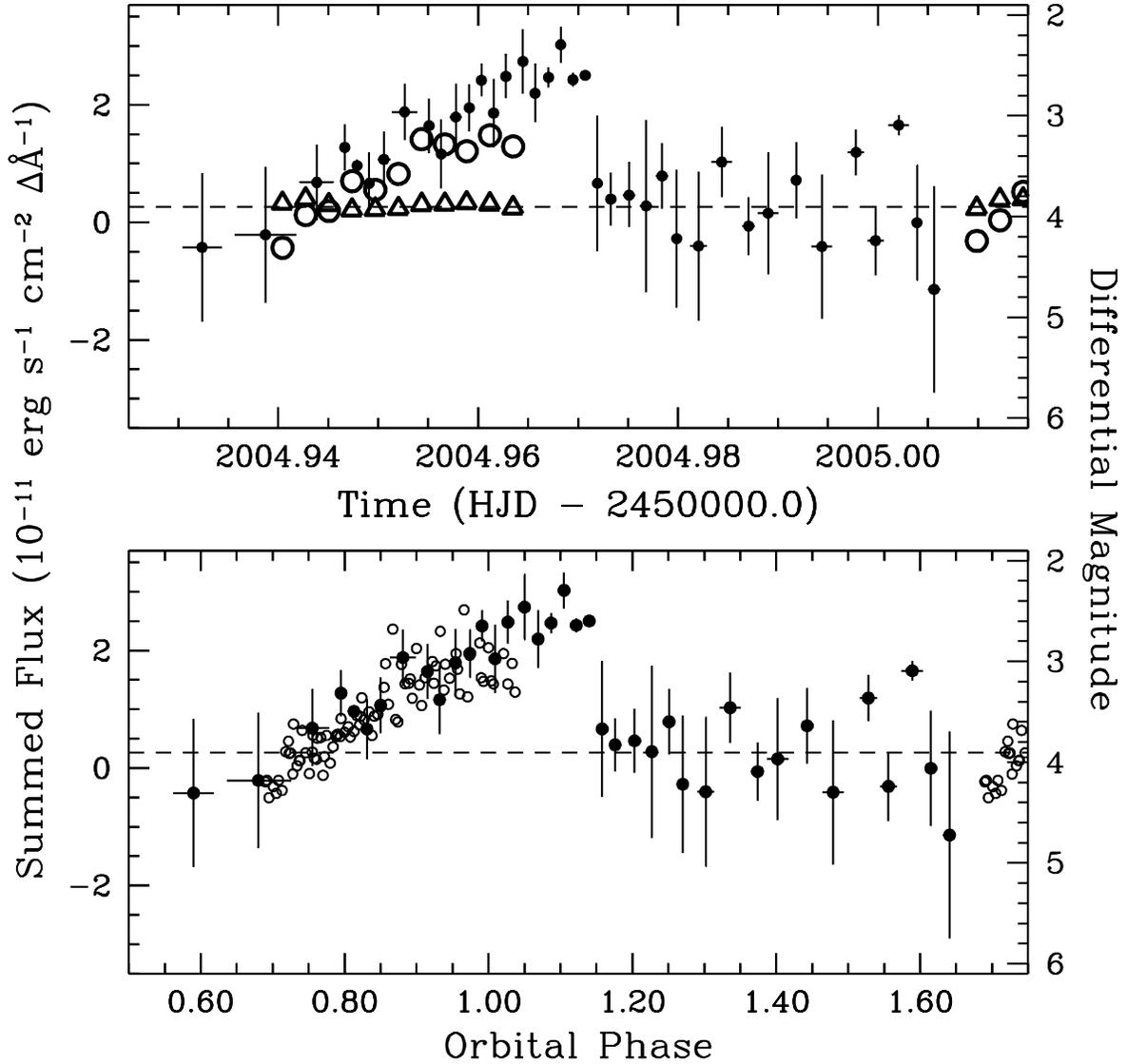}
\epsscale{1.00}
\figcaption{Optical and FUV light curves of VV Pup.  The optical 
data are presented as in Figure \ref{f-optlc}.  In the top panel, 
the FUV continuum (large unfilled circles) and emission (large 
unfilled triangles) flux data are overplotted for the {\em FUSE} 
visits that exactly overlap the time range of the optical 
observations.  In the bottom panel, the FUV continuum flux data 
(small unfilled circles), folded on the ephemeris of 
\citet{walker65}, are overplotted.  The vertical axes on the 
left in both panels give the flux scale; 
the vertical axes on the right in both panels give the optical 
magnitude scale.  The range of the flux axis has been chosen so 
that the phase-folded continuum flux points most closely overlap 
the optical data (assuming that the start of the continuum flux 
data corresponds to the ``zero'' level of the optical light curve).  
\label{f-lccomp}}
\end{figure}

Figure \ref{f-lccomp} shows a comparison of the FUV and 
optical light curves of VV Pup.  
The top panel shows our optical data 
as a function of HJD; we have overplotted the {\em FUSE} data 
that exactly match the time of the optical observations.  
It is already clear that the optical and FUV continuum data 
behave similarly, while the FUV emission data display substantially
different behavior from the optical data.
The bottom panel in the figure shows the optical  
and FUV continuum data folded on the ephemeris 
of \citet{walker65}.  There is good agreement between the light curve
shapes of the folded optical and FUV continuum data.

As mentioned above, our optical light curve is very similar in 
phasing and overall appearance to the optical light curve of 
VV Pup in \citet{patterson84}:\
it begins to rise from the ``zero'' level at an approximately 
constant rate at $\phi\approx0.70$, reaches maximum at $\phi\approx0.05$, 
levels off until $\phi\approx0.15$, and then drops steeply back 
to the zero level in only a few hundredths of an orbital cycle.
The FUV continuum light curve, however, appears to peak and 
already begin declining well before the optical peak, and also well 
before the peak found in a light curve obtained from the 
{\em Extreme Ultraviolet Explorer} ({\em EUVE\/}) satellite 
\citep[][their Figure 2]{vennes95}.  It is possible that this 
is caused by the accretion stream occultation feature suggested 
by \citet{vennes95}, but we cannot make a firm conclusion without 
having {\em FUSE} data from later orbital phases.

\subsection{Far-ultraviolet Spectrum}
\label{s-fuvspec}

\begin{figure}[tb]
\epsscale{0.90}
\plotone{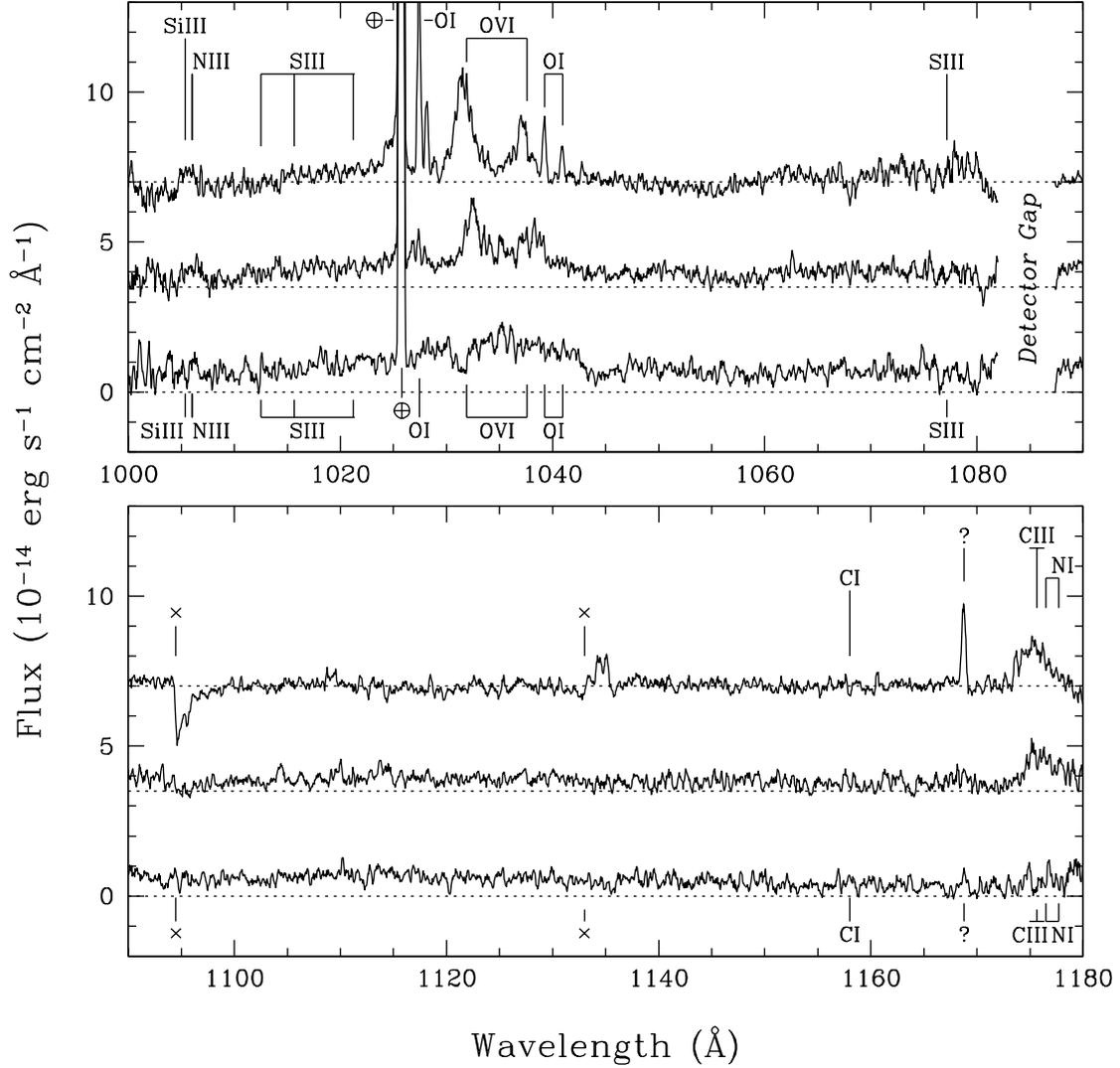}
\epsscale{1.00}
\figcaption{Time-resolved FUV spectra of VV Pup.  The top panel 
shows the wavelength range 1000--1090 \AA, the bottom panel shows 
1090--1180 \AA.  
The spectra were rebinned to a uniform dispersion of 
0.05 \AA\ pixel$^{-1}$ during processing, and have been boxcar 
smoothed by 5 pixels in this figure. 
The three spectra correspond to phase ranges of 
$\Delta\phi_{1} = 0.67$--0.79 (upper spectrum in each panel), 
$\Delta\phi_{2} = 0.80$--0.89 (middle), and 
$\Delta\phi_{3} = 0.90$--0.06 (bottom).
The bottom spectrum in each panel shows the true flux level, 
while the other spectra have been successively offset by 
$+2.5\times10^{-14}$ erg s$^{-1}$ cm$^{-2}$ \AA$^{-1}$ (the 
dotted lines show the true zero level for each spectrum).
A number of line transitions have been identified and marked 
with vertical bars; multiplet transitions of the same ion are 
joined by a horizontal bar.  
The features marked ``$\times$'' may be artifacts, since they 
are located near the ends of detector segments.  
The feature at 1168 \AA\ marked ``?'' possibly corresponds 
to \ion{He}{1} $\lambda584$ in second order \citep{feldman01}.
\label{f-fuvspec}}
\end{figure}

We used the procedure described in Section \ref{s-fuvlc} and the 
ephemeris of \citet{walker65} to extract three additional 
time-resolved spectra from the raw {\em FUSE} data.  
Data from each of the 11 {\em FUSE} visits corresponding to 
orbital phase ranges of $\Delta\phi_{1} = 0.67$--0.79 (total 
exposure time of 7069 s), $\Delta\phi_{2} = 0.80$--0.89 (5442 s), 
and $\Delta\phi_{3} = 0.90$--0.06 (5929 s) were combined to 
produce these spectra (see Figure \ref{f-fuvspec}).  They have 
weak, mostly flat continua with a ``hump'' between 
$\approx$1010--1050 \AA. The spectra lack prominent emission 
or absorption features with the exception of several \ion{O}{1} 
and \ion{O}{6} emission lines around 1030 \AA\ (identified in 
Section \ref{s-fuvlc}) and a \ion{C}{3} emission feature at 
1175 \AA.  The wavelengths of a few lines of other ions are 
identified in Figure \ref{f-fuvspec}, but they are not entirely 
convincing as real features.

As is reflected in the FUV light curve of VV Pup (see 
Section \ref{s-fuvlc}), the continuum flux level in the spectra 
increases at larger orbital phases.   Despite significant changes 
in the shapes of the \ion{O}{6} emission line profiles (it is 
arguable whether these lines are even still present in the 
$\Delta\phi_{3}$ spectrum), the total flux in this wavelength 
region remains relatively constant (compared to the continuum 
flux change -- see Figure \ref{f-fuvlc}).  
The narrow \ion{O}{1} 
lines seen in the $\Delta\phi_{1}$ spectrum are probably airglow 
features, since approximately 65\% of the combined data for this 
spectrum were obtained during spacecraft day (all of the 
$\Delta\phi_{2}$ and $\Delta\phi_{3}$ data, which do not show 
the \ion{O}{1} emission, were obtained during spacecraft night).  
The prominence of the \ion{O}{6} features in the spectrum of VV Pup 
may be related to the possible oxygen overabundance suggested 
by \citet{vennes95} from their EUV data.  On the other hand, a 
recent {\em FUSE} spectrum of the prototype magnetic CV, AM Her, 
also shows strong \ion{O}{6} emission \citep{hutchings02}, so this 
may be a common FUV feature of magnetic CVs (or CVs in general).
We do not observe narrow emission components in the \ion{O}{6}
profiles of VV Pup, as seen by \citet{mauche99} in 
{\em ORFEUS II} FUV spectra 
of the polar EX Hydrae, and by \citet{mauche98} and 
\citet{hutchings02} in {\em ORFEUS II} and {\em FUSE} spectra, 
respectively, of AM Her.  In the case of EX Hya, the narrow 
component velocities suggest an origin on the WD, whereas in 
AM Her, the velocities suggest an origin on the irradiated 
inner face of the secondary star.  Given that these narrow 
emission components can originate from very different regions 
in these two polars, it is not so surprising that they might be 
absent altogether in a third polar.

\section{Discussion}

\subsection{System Geometry}

We can use the geometric model for VV Pup presented by 
\citet{patterson84} to estimate which system components are visible 
along the lines-of-sight represented by each of our phase-resolved 
{\em FUSE} spectra.  The phase interval $\Delta\phi_{1}$ starts 
shortly after the superior conjunction of the secondary star 
(signaled by a photometric infrared minimum) at $\phi\approx0.6$.  
Optical circular polarization appears at $\phi\approx0.75$ as the accretion 
spot on the WD first becomes visible.  The accretion funnel is 
visible earlier, possibly for the entire $\Delta\phi_{1}$ interval.  
The broad \ion{O}{6} emission lines present in the $\Delta\phi_{1}$ 
spectrum display little or no Doppler shift, and likely originate 
in the accretion funnel, which is directed approximately 
perpendicular to the observer's line-of-sight during most of this 
phase interval.  These are high excitation lines and, hence, most 
likely are formed in the hot portion of the funnel, close to the WD 
surface (henceforth, we refer to this as the ``inner funnel'' to 
signify its proximity to the WD in the CV system).

The phase interval $\Delta\phi_{2}$ brackets the phase of maximum 
elongation, when the WD (secondary star) is maximally receding from 
(approaching) the observer.  The accretion funnel is now viewed in 
recession from the observer, due to both the binary motion of the WD 
and the matter flow motion within the funnel.  This motion is 
reflected in the redward shift of the \ion{O}{6} line peaks 
compared to the previous spectrum.  In addition, the oxygen lines 
are weaker than in the $\Delta\phi_{1}$ spectrum, possibly because 
the cooler outer part of the accretion funnel is partially obscuring 
the hotter part closer to the WD where the \ion{O}{6} lines 
presumably originate (however, see the discussion below).

The phase of maximum optical/UV light occurs in $\Delta\phi_{3}$ 
at $\phi\approx0.98$, when the magnetic pole of the WD and associated 
accretion spot (located at an azimuth of $\psi\approx50^{\circ}$; 
\citealt{cropper88}) rotates fully and directly into view.  We 
would, thus, expect emission from the high velocity inner end of the 
accretion funnel to have maximum redshift in this phase interval.  
Unfortunately, the \ion{O}{6} emission lines appear to have 
essentially disappeared in the phase-resolved spectrum.  
The inclination 
($i\approx75^{\circ}$) and magnetic pole colatitude 
($\delta\approx150^{\circ}$) in VV Pup (\citealt{cropper88}) make 
it unlikely that this is the result of complete obscuration 
of the emission region in the hot inner funnel by another 
part of the funnel, unless the obscuring funnel material is 
located far from the WD surface (see discussion in 
\citealt{patterson84}, their Section IIIb).  
Additionally, this would only be expected to affect the system 
for a short time (e.g., to produce the narrow dip seen 
by \citealt{vennes95} in the EUV light curve of VV Pup).  
The FUV emission light curve (see Figure \ref{f-fuvlc})
shows that the flux in the wavelength region around the oxygen lines
increases during $\Delta\phi_{3}$, contrary to our expectation 
if the emission lines were suddenly blocked from view during 
this orbital phase interval.
In addition, the $\Delta\phi_{3}$ spectrum displays a broad hump 
where the \ion{O}{6} line peaks were located in the 
$\Delta\phi_{2}$ spectrum.  The red edge of this hump is located 
considerably redward (at $\approx1043$ \AA) of the red edge of 
the \ion{O}{6} $\lambda1037.6$ line in the $\Delta\phi_{2}$ 
spectrum (at $\approx1039$ \AA).  
A similar effect was seen by \citet{ferrario99} in the emission 
lines of model optical spectra for a high inclination polar near 
the phase of maximum redshift (e.g., their Figure 10).  They 
attributed it to a combination of velocity smearing and 
self-obscuration in the funnel.
Thus, we suggest that the \ion{O}{6} 
emission lines are still present and visible in the 
$\Delta\phi_{3}$ spectrum of VV Pup, but they have been both substantially 
redshifted and severely broadened, to the point of being 
unrecognizable as peaked emission line profiles, by our ``deepest'' 
view down the accretion funnel to its inner, high velocity end.

\subsection{Accretion Region Characteristics}

\begin{figure}[tb]
\epsscale{1.00}
\plotone{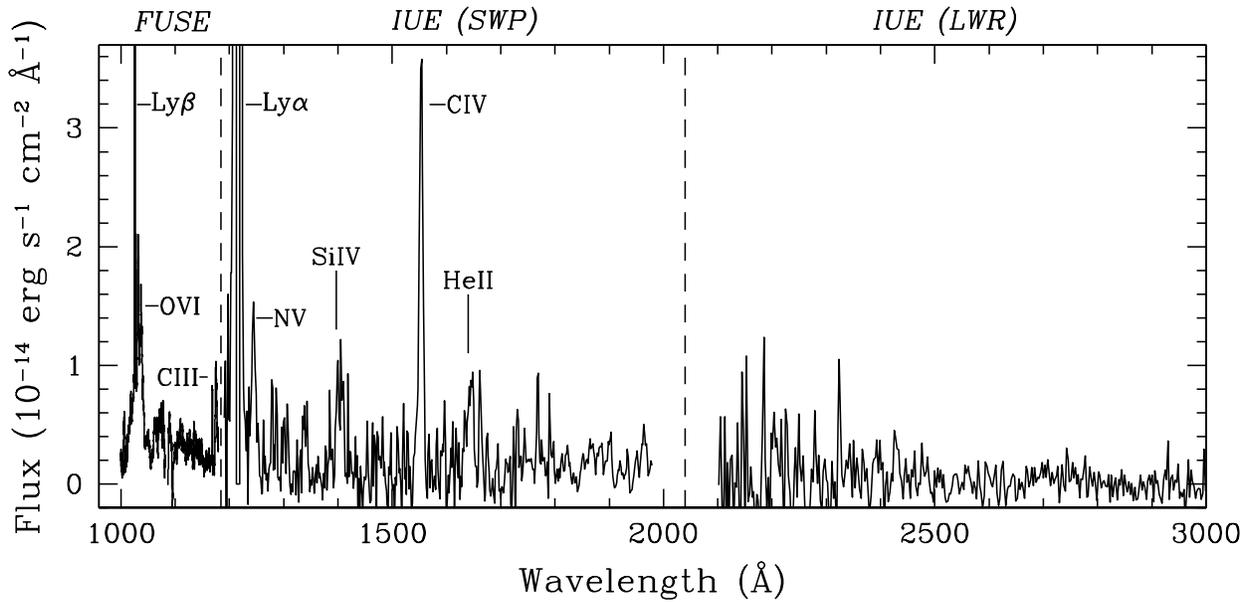}
\epsscale{1.00}
\figcaption{Combined {\em FUSE} and {\em IUE} spectra of VV Pup.  
The {\em IUE} spectra are SWP07868 (14,400 s obtained on 1980 
February 04) and LWR07309 (7200 s obtained on 1980 March 26).  
Vertical dashed lines mark the boundaries between the {\em FUSE}, 
{\em IUE\ } SWP, and {\em IUE\ } LWR wavelength regions.  Several 
spectral features are indicated.
\label{f-iuespec}}
\end{figure}

VV Pup was observed several times with the {\em International 
Ultraviolet Explorer} ({\em IUE\/}) satellite 
\citep[e.g.,][]{patterson84}.  We extracted the available data 
from the {\em IUE} archive, and have plotted a representative 
short wavelength spectrum (SWP07868, 14,400 s obtained on 
1980 February 04) and the only long wavelength spectrum 
(LWR07309, 7200 s obtained on 1980 March 26) in 
Figure \ref{f-iuespec}.  A mean FUV spectrum constructed by 
combining our three phase-resolved spectra of VV Pup is also 
shown in the figure.  The flux levels of all three satellite 
spectrum segments agree well at their boundaries (despite the 
fact that the {\em IUE} spectra are each averaged over several 
orbital cycles of the CV, whereas the combined {\em FUSE} spectrum 
covers only $\Delta\phi\approx0.40$ during the brightest portion 
of the orbit). 

\begin{figure}[tb]
\epsscale{0.95}
\plotone{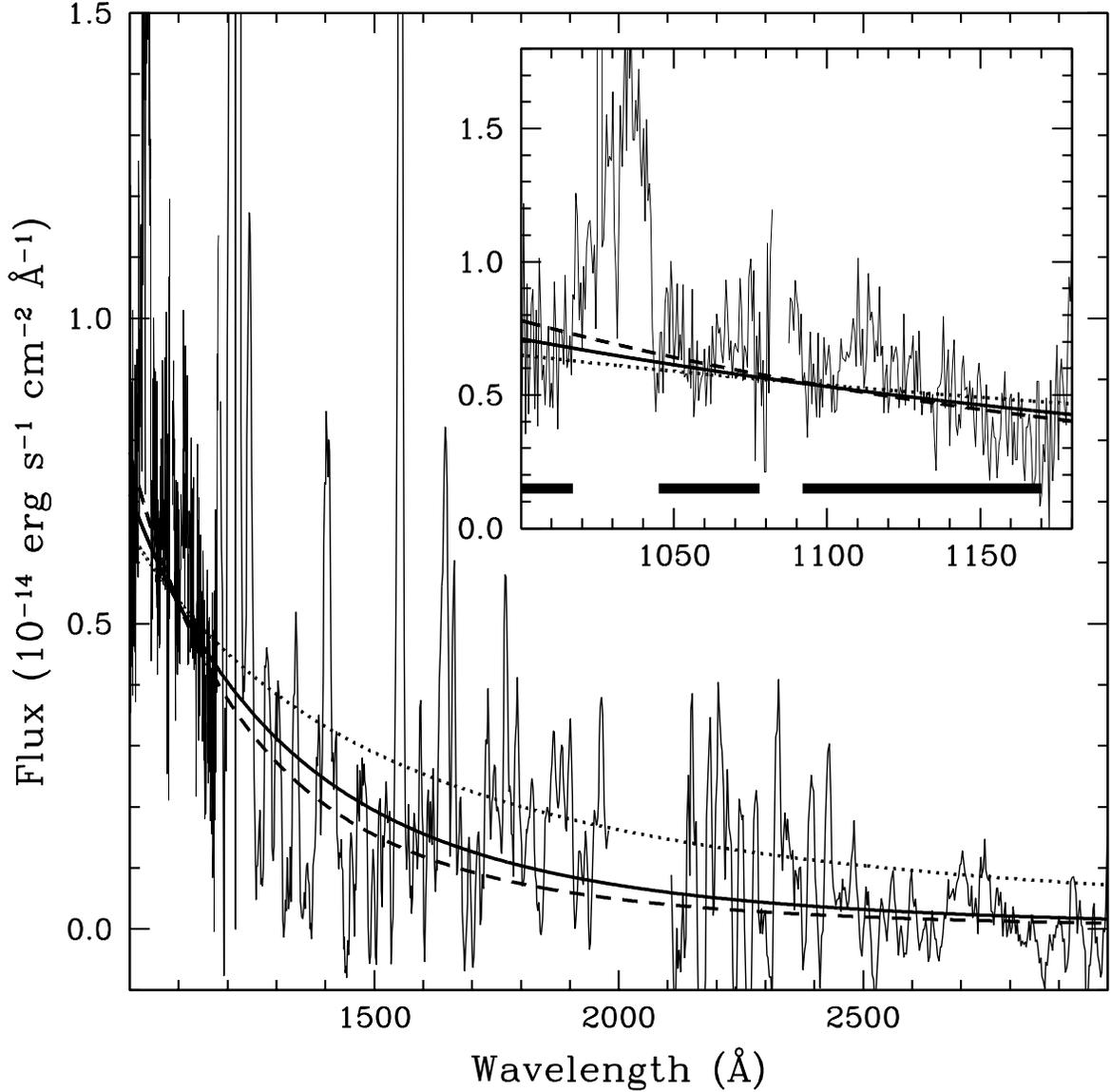}
\epsscale{0.95}
\figcaption{Main panel shows the $\Delta\phi_{3} = 0.90$--0.06 
spectrum from Figure \ref{f-fuvspec} (rebinned to a dispersion of 
0.5 \AA\ pix$^{-1}$) and the {\em IUE} spectra of VV Pup from 
Figure \ref{f-iuespec} (faint solid line) with three representative 
models:\ a power law of the form $F_{\lambda}\propto\lambda^{\alpha}$ 
where $\alpha=-2$ (dotted line) or $-4$ (dashed line), and a 
blackbody function with $T_{\rm bb}=90,000$ K (dark solid line). 
The inset panel shows a close-up of only the {\em FUSE} spectrum 
with the same three models.  The horizontal bars show the wavelength 
ranges used for determining the scaling factors that best fit each 
model to the {\em FUSE} data (i.e., the {\em IUE} data were {\em not} 
used to normalize the models).  Bright airglow lines 
(and the \ion{C}{4} $\lambda1550$ emission line) are 
truncated in both plots.  The {\em IUE} spectra have been 
boxcar-smoothed by 5 pixels.
\label{f-specfit}}
\end{figure}

In Figure \ref{f-specfit}, we have overplotted three simple models 
onto a {\em FUSE\/}+{\em IUE} spectrum of VV Pup.  The same 
{\em IUE} data from Figure \ref{f-iuespec} are used, but the 
$\Delta\phi_{3}$ (i.e., strongest continuum) {\em FUSE} spectrum 
is used instead of the mean spectrum.  The models are:\ a power 
law of the form $F_{\lambda}\propto\lambda^{\alpha}$ where 
$\alpha=-2$ or $-4$, and a blackbody function with 
$T_{\rm bb}=90,000$ K.  
The inset panel of the figure shows the 
models with only the {\em FUSE} data; the horizontal bars mark the 
wavelength regions used to normalize the three models.  In the 
case of the power law, the exponent was kept fixed and the 
multiplicative scaling factor that minimized 
the r.m.s.\ deviation between the 
data and the fit was determined.  A similar process was followed 
for the blackbody fit:\ the normalization with minimum r.m.s.\ was 
determined for fixed temperatures in 5000 K steps in the range 
50,000--500,000 K, and we kept the model with the overall 
smallest r.m.s.\ deviation.  (The r.m.s.\ values increase smoothly 
on either side of the model with $T_{\rm bb}=90,000$ K.)  All 
three of these models provide adequate fits to the continuum 
shape of the phase-resolved {\em FUSE} spectrum within the 
relatively loose limits allowed by its low S/N.
By comparison, a 9000 K WD \citep{liebert78} would contribute very 
little flux in the FUV (a factor of $\approx6\times10^{-5}$ 
smaller than the 90,000 K blackbody at 1100 \AA).
The {\em IUE} data were not used when we determined the model 
normalizations.  Yet, the $\alpha=-4$ power law and hot blackbody 
both still provide good fits to the data when extended into the 
{\em IUE} spectra.
Assuming that the normalization to the {\em FUSE} data is correct, 
then the {\em IUE} data effectively rule out the $\alpha=-2$ power 
law (equivalent to cooler temperatures) as a good fit to the UV 
spectral energy distribution of VV Pup.  

\citet{vennes95} estimated an accretion region temperature of 
300,000 K from WD synthetic spectrum fits to their {\em EUVE} 
spectrum of VV Pup.  Although hotter than the nominal temperature 
found from a blackbody fit to our {\em FUSE} data, the {\em EUVE} 
result is not inconsistent with the {\em FUSE} data -- a 300,000 K 
blackbody would produce a curve almost indistinguishable from 
the $\alpha=-4$ power law.  We expect to find a lower temperature 
in our data since we have moved to longer wavelengths, and are 
presumably observing cooler zones of a more extended accretion 
region than is detectable in the EUV (the {\em EUVE} spectrum 
covers $\approx80$--140 \AA\ compared to $\approx1000$--1180 \AA\ 
for {\em FUSE}).  
\citet{boris97} found a range of accretion spot 
temperatures of $\approx25,000$--$40,000$ K in a sample of seven 
AM Her systems using ultraviolet (1200--3000 \AA) data from {\em IUE}. 
Thus, our best model temperature of 90,000 K lies between the 
temperatures derived from longer wavelength {\em IUE} data and 
shorter wavelength {\em EUVE} data, as expected.

The flux of our 90,000 K blackbody model at $\lambda=5500$ \AA\ 
is $F_{\rm 5500}\approx1.7\times10^{-17}$ erg s$^{-1}$ cm$^{-2}$ \AA$^{-1}$.
Using $F_{\rm 5500}\approx3.7\times10^{-9}$ erg s$^{-1}$ cm$^{-2}$ \AA$^{-1}$ 
for a 0th magnitude AOV star \citep{colina96}, then the magnitude 
of our 90,000 K blackbody model at $\lambda=5500$ \AA\ would 
be $\approx 20.8$, much fainter than the optical magnitude 
of $\approx 16.5$ determined for VV Pup from our ground-based 
data (see Section \ref{s-optlc}).
In this regard, it is unclear why the optical and FUV--EUV light 
curves of VV Pup are so similar in appearance, since the dominant 
source of FUV flux does not contribute significantly at optical 
wavelengths.  Reprocessing of the FUV radiation in material 
around the accretion region may play a role in this.

The normalization factor ($N = 4 \pi r^2/d^2 = 1.13\times10^{-25}$) 
determined for the 90,000 K blackbody model corresponds 
to a circular emitting region with radius given by
     \begin{equation}
     r_{\rm acc} \approx (293 \,\, {\rm km})(d/100 \,\, {\rm pc}),
     \end{equation}
where $d$ is the distance to VV Pup in pc.  This gives 
$r_{\rm acc}\approx425$ km for $d=145$ pc \citep{bailey81,vennes95}.  We would 
expect that the detectable accretion region should be somewhat larger 
in the FUV than in the EUV (because we are able to detect radiation 
from cooler regions), and this estimate is $\approx4$ times the 
$r_{\rm acc,EUV}=110$ km obtained by \citet{vennes95} from timing 
and geometric considerations using a 3-d hemispherical spot model 
(unfortunately, we lack the necessary timing information to repeat 
their 3-d calculation because of the incomplete coverage of the 
bright hump in our FUV light curve).  A hotter blackbody model 
would require a larger normalization factor (thus yielding a 
smaller $r_{\rm acc}$ for the same distance); for example, if 
$T_{\rm bb}=300,000$ K, then $N=2.7\times10^{-26}$ and 
$r_{\rm acc}\approx210$ km.
By comparison, the 9000 K WD found by \citet{liebert78} would need 
a radius of almost 55,000 km to account for the observed FUV flux 
of VV Pup at a distance of 145 pc (or be located at a distance of 
only $\approx15$ pc if the WD radius is a more reasonable 6000 km).

\section{Conclusions}

Analysis of our {\em FUSE} observations of VV Pup has revealed 
many of the FUV characteristics of this magnetic CV.  
The shape of the FUV light curve is consistent with that seen 
at other wavelengths (optical, EUV, X-ray).
Likewise, the characteristics of the phase-resolved FUV spectra 
during the bright portion of VV Pup's orbit (when the accreting 
magnetic pole is visible) agree with expectations from the 
geometric model for this system presented by \citet{patterson84}.
These include an increase in the continuum level at later orbital 
phases (as the magnetic pole and its associated accretion region rotate 
into more direct view) and changes in the shape and velocity 
shift of the prominent \ion{O}{6} emission lines that we 
interpret in the context of the lines originating in the high 
velocity, hot, inner region of the accretion funnel located 
close to the WD surface.
A hot temperature ($T\gtrsim90,000$ K) is favored for a simple 
model of the FUV spectral energy distribution in VV Pup, in terms 
of both the fit to the {\em FUSE} and archival {\em IUE} continuum 
in the wavelength range 1000--3000 \AA, and the size of the accretion 
region estimated from the normalization of the model.  
Although we expect {\em a priori} that observations in the FUV 
should detect a cooler (and more extended) accretion region than 
did the {\em EUVE} observations of \citet{vennes95}, we cannot 
rule out a temperature as high as the 300,000 K determined for 
the EUV as an adequate fit to the {\em FUSE} data.

\acknowledgements

We thank Paul Barrett for his input regarding the {\em FUSE} 
spectrum of VV Pup.  This work was supported by NASA {\em FUSE} 
grant NAG 5-10343.  B.\ T.\ G. was supported by a PPARC Advanced Fellowship.
We made use of the SIMBAD database, operated 
at CDS, Strasbourg, France.

%%% BIBLIOGRAPHY

%%% TABLES
\begin{deluxetable}{cccc}
\tabletypesize{\small}
\tablewidth{0pt}
\tablecaption{{\em FUSE} Observation Log \label{t-fuselog}}
\tablehead{
\colhead{Visit} &
\colhead{Start Time} &
\colhead{Total Exposure} &
\colhead{$\phi$\tablenotemark{a}} \\
\colhead{ } &
\colhead{(HJD$-$2450000)} &
\colhead{(s)} &
\colhead{ } 
}
\startdata
 1 & 2004.52312 & \phn501 & 0.721 \\
 2 & 2004.59228 & \phn741 & 0.712 \\
 3 & 2004.66153 & \phn981 & 0.705 \\
 4 & 2004.73099 &    1310 & 0.701 \\
 5 & 2004.80044 &    1770 & 0.697 \\
 6 & 2004.86983 &    2180 & 0.692 \\
 7 & 2004.93937 &    2270 & 0.689 \\
 8 & 2005.00881 &    2270 & 0.684 \\
 9 & 2005.07812 &    2270 & 0.678 \\
10 & 2005.14751 &    2271 & 0.673 \\
11 & 2005.21740 &    2014 & 0.675
\enddata 
\tablenotetext{a}{Orbital phase at start of exposure 
from the ephemeris of \citet{walker65}.}
\end{deluxetable}

%%% FIGURES
%%%%%%%%%%%%%%%%%%%%%%%%%%%%%%%%%%%%%%%%%%%%%%%%%%%%%%%%%%%%%%%%%%%
%%BoundingBox: 18 155 570 695
%%%%%%%%%%%%%%%%%%%%%%%%%%%%%%%%%%%%%%%%%%%%%%%%%%%%%%%%%%%%%%%%%%%
%%%%%%%%%%%%%%%%%%%%%%%%%%%%%%%%%%%%%%%%%%%%%%%%%%%%%%%%%%%%%%%%%%%
%%BoundingBox: 15 144 570 450
%%%%%%%%%%%%%%%%%%%%%%%%%%%%%%%%%%%%%%%%%%%%%%%%%%%%%%%%%%%%%%%%%%%
%%%%%%%%%%%%%%%%%%%%%%%%%%%%%%%%%%%%%%%%%%%%%%%%%%%%%%%%%%%%%%%%%%%
%%BoundingBox: 18 160 575 695
%%%%%%%%%%%%%%%%%%%%%%%%%%%%%%%%%%%%%%%%%%%%%%%%%%%%%%%%%%%%%%%%%%%
%%%%%%%%%%%%%%%%%%%%%%%%%%%%%%%%%%%%%%%%%%%%%%%%%%%%%%%%%%%%%%%%%%%
%%BoundingBox: 20 150 580 695
%%%%%%%%%%%%%%%%%%%%%%%%%%%%%%%%%%%%%%%%%%%%%%%%%%%%%%%%%%%%%%%%%%%
%%%%%%%%%%%%%%%%%%%%%%%%%%%%%%%%%%%%%%%%%%%%%%%%%%%%%%%%%%%%%%%%%%%
%%BoundingBox: 40 165 580 425
%%%%%%%%%%%%%%%%%%%%%%%%%%%%%%%%%%%%%%%%%%%%%%%%%%%%%%%%%%%%%%%%%%%
%%BoundingBox: 18 155 570 700
%%%%%%%%%%%%%%%%%%%%%%%%%%%%%%%%%%%%%%%%%%%%%%%%%%%%%%%%%%%%%%%%%%%


\begin{thebibliography}{}

\bibitem[Bailey(1981)]{bailey81} Bailey, J. 1981, \mnras, 197, 31 

\bibitem[Barrett \& Chanmugam(1985)]{barrett85} Barrett, P.\ E., 
Chanmugam, G. 1985, \apj, 298, 743

\bibitem[Colina, Bohlin, \& Castelli(1996)]{colina96} Colina, L., 
Bohlin, R., Castelli, F. 1996, Space Telescope Science Institute 
Observatory Support Group Calibration Report, 
OSG-CAL-96-01\footnote{See 
\url{http://www.stsci.edu/instruments/observatory/reports.html}.}

\bibitem[Cropper(1988)]{cropper88} Cropper, M. 1988, \mnras, 231, 597

\bibitem[Downes et al.(2001)]{downes01} Downes, R.\ A., Webbink, 
R.\ F., Shara, M.\ M., Ritter, H., Kolb, U., Duerbeck, H.\ W. 2001, 
\pasp, 113, 764 

\bibitem[Feldman et al.(2001)]{feldman01} Feldman, P.\ D., Sahnow, 
D.\ J., Kruk, J.\ W., Murphy, E.\ M., Moos, H.\ W. 2001, 
J.\ Geophys.\ Res., 106, 8119

\bibitem[Ferrario \& Wehrse(1999)]{ferrario99} Ferrario, L., Wehrse, R. 
1999, \mnras, 310, 189 

\bibitem[G\"{a}nsicke(1997)]{boris97} G\"{a}nsicke, B.\ T. 1997, 
PhD thesis (U.\ of G\"{o}ttingen)

\bibitem[Herbig(1960)]{herbig60} Herbig, G.\ H. 1960, \apj, 132, 76

\bibitem[Hessman, G\"{a}nsicke, \& Mattei(2000)]{hessman00} Hessman, 
F.\ V., G\"{a}nsicke, B.\ T., Mattei, J.\ A. 2000, \aap, 361, 952

\bibitem[Hutchings et al.(2002)]{hutchings02} Hutchings, J.\ B., 
Fullerton, A.\ W., Cowley, A.\ P., Schmidtke, P.\ C. 2002, \aj, 123, 2841

\bibitem[Imamura, Steiman-Cameron, \& Wolff(2000)]{imamura00} Imamura, 
J.\ N., Steiman-Cameron, T.\ Y., Wolff, M.\ T. 2000, \pasp, 112, 18

\bibitem[Liebert \& Stockman(1979)]{liebert79} Liebert, J., Stockman, 
H.\ S. 1979, \apj, 229, 652

\bibitem[Liebert et al.(1978)]{liebert78} Liebert, J., Stockman, H.\ S., 
Angel, J.\ R.\ P., Woolf, N.\ J., Hege, K., Margon, B. 1978, \apj, 225, 201

\bibitem[Mauche(1999)]{mauche99} Mauche, C.\ W. 1999, \apj, 520, 822

\bibitem[Mauche \& Raymond(1998)]{mauche98} Mauche, C.\ W., Raymond, J.\ C. 1998, \apj, 505, 869

\bibitem[Patterson et al.(1984)]{patterson84} Patterson, J., Beuermann, 
K., Lamb, D.\ Q., Fabbiano, G., Raymond, J.\ C., Swank, J., White, 
N.\ E. 1984, \apj, 279, 785

%%% the following reference has 35 authors
\bibitem[Sahnow et al.(2000)]{sahnow00} Sahnow, D.\ J., et al. 2000, 
Proc.\ SPIE, 4013, 334

\bibitem[Schmidt et al.(1999)]{schmidt99} Schmidt, G.\ D., Hoard, D.\ W., 
Szkody, P., Melia, F., Honeycutt, R.\ K., Wagner, R.\ M. 1999, \apj, 525, 407

\bibitem[Szkody, Bailey, \& Hough(1983)]{szkody83} Szkody, P., Bailey, 
J.\ A., Hough, J.\ H. 1983, \mnras, 203, 749

\bibitem[Szkody \& Capps(1980)]{szkody80} Szkody, P., Capps, R.\ W. 1980, 
\aj, 85, 882

\bibitem[Tapia(1977)]{tapia77} Tapia, S. 1977, \iaucirc, 3054, 1

\bibitem[van Gent(1931)]{vangent31} van Gent, H. 1931, \bain, 6, 93

\bibitem[Vennes et al.(1995)]{vennes95} Vennes, S., Szkody, P., Sion, 
E.\ M., Long, K.\ S. 1995, \apj, 445, 921

\bibitem[Visvanathan \& Wickramasinghe(1979)]{visv79} Visvanathan, N., 
Wickramasinghe, D.\ T. 1979, \nat, 281, 47

\bibitem[Walker(1965)]{walker65} Walker, M.\ F. 1965, Comm.\ Konkoly Obs., 
57, 1

\bibitem[Warner(1995)]{warner95}Warner, B. 1995, Cataclysmic Variable 
Stars (Cambridge: Cambridge University Press)

\bibitem[Warner \& Nather(1972)]{warner72} Warner, B., Nather, R.\ E. 1972, 
\mnras, 156, 305

\bibitem[Wickramasinghe, Ferrario, \& Bailey(1989)]{wick89} Wickramasinghe, 
D.\ T., Ferrario, L., Bailey, J. 1989, \apjl, 342, L35

\bibitem[Wickramasinghe \& Visvanathan(1979)]{wick79} Wickramasinghe, 
D.\ T., Visvanathan, N. 1979, Proc.\ Astron. Soc.\ Australia, 3, 311

\end{thebibliography}
\end{document}